# Estimating the Environmental Cost of Shrimp Farming in Coastal Areas of Chittagong and Cox's bazaar in Bangladesh

Mohammad Nur Nobi[a] and Dr. A N M Moinul Islam[b]

**Abstract:**

During the last three decades, shrimp has remained one of the major export items in Bangladesh. It contributes to the development of this country by enhancing export earnings and promoting employment. However, coastal wetlands and agricultural lands are used for shrimp culture, which reduces agricultural opportunity and peasants' income, and destroys the mangroves and coastal eco-system. These are the external environmental costs that are not reflected in farmers' price and output decisions. This study has aimed to estimate those external environmental costs through the contingent valuation method. The calculated environmental cost of shrimp farming is $13.66 per acre per year. Findings suggest that current shrimp production and shrimp price will no longer be optimal once the external costs are internalized. Thus alternative policy recommendations have been proposed so that shrimp farming becomes a sustainable and equitable means of aquaculture.

a. Department of Economics, University of Chittagong, Bangladesh, Email: nurnobi@cu.ac.bd.
b. Asian University for Women, Chittagong, Bangladesh.

# Estimating the Environmental Cost of Shrimp Farming in Coastal Areas of Chittagong and Cox's bazaar in Bangladesh

1. Introduction

Shrimp has remained a dominant export item in Bangladesh since 1980. At present, Bangladesh produces 2.5% of the entire international shrimp production regime (Anwar, 2003). In 2011, the total production of marine shrimp in Bangladesh was approximately 225 thousand million tons (Jory and Cabrera, 2012), and in FY 2010-2011, its contribution was 2.73% of export earnings in the economy (DoF). Though shrimp farming is profitable to the coastal community and the country, random and disheveled farming is assumed to harm the environment and influence residents' environmental concerns (Alam and Hossain, 2018) in those areas. Shrimp culture is a labor-intensive (low capital intensive) production. Both extensive and semi-extensive shrimp farming is in practice in Bangladesh with labor intensity. So, the impact of shrimp farming starts from the methods used in collecting the shrimp fry, which destroys other marine species, and the cultivation of shrimp reduces agricultural land and production. During the last twenty years, there was a dramatic decline in live stocks with 21% due to the reduction of grazing fields (Anwar, 2003) in the coastal areas in Bangladesh.

Shrimp farming craves a distinct process that forces farmers to dig ponds in the coastal area using saline water. So, it has two effects in the coastal area, and the former reduces coastal wetland by digging ponds and draining saline water inside the coastal area, which hinders the eco-system and has a spillover effect on agricultural production. The rapid increase of salinity in the coastal belt reduces the fertility of agricultural land. The coastal areas of Shatkhira, Khulna, Bagherhat, and Cox's bazaar are mainly shrimp-producing zones in Bangladesh. Annual shrimp & prawn production is 28,514 Mt (headless) which earns 16,150 Million BDT as foreign exchange.

Since it is profitable, the landowners then migrate their cultivable land into aquaculture. The following graph (Graph 1) also proves that the land employed in shrimp culture increases in Cox's Bazar district though there is a decreasing trend in Chittagong district. The primary input of shrimp production is the prawn fry dependent on wild postlarvae (PL), which is now also produced in hatcheries. However, the quality of hatchery PL is questionable since there is no regulation though 42 hatcheries are in operation out of 81 in Bangladesh (Ahmed, 2008).

Table 1: District-wise shrimp production for FY 2012-2013

| District | Area(Ha) | | | Shrimp and Prawn Production(Mt) | | | | Other Fish | Total |
|---|---|---|---|---|---|---|---|---|---|
| | Baghda | Galda | Total | Baghda | Galda | Other shrimp /Prawn | Total shrimp/ Prawn | | |
| Chittagong | 2805.0 | 468.72 | 3273.72 | 838.00 | 218.24 | 30.87 | 1087.11 | 1087.11 | 2460.86 |
| Cox's Bazar | 43377.0 | 330.20 | 43707.2 | 12560.0 | 122.94 | 516.50 | 13199.44 | 6330.72 | 19530.16 |

*Source: Statistical Year Book 2012-2013*

In the early 1980s, the primary source of shrimp farming was wild fry collected from coastal and tidal waters. About 0.45 million people, including women, children, and landless, are engaged in the fry collecting section (FAO, Web1, and DoF, 2003).

Shrimp farming expanded rapidly without careful planning and a strategic system despite the absence of authorized control or national strategy in Bangladesh. Thus it increased from 20,000 ha in 1980 to 70,000 ha in 1985; 115,000 ha in 1989, 203071 at 2003-2004 FY, and 275274 ha at 2012-2013 FY. As a result, expansion of shrimp farming/aquaculture depletes mangrove forests by a much higher rate, and 50% of 18,200 ha of mangrove in the coastal area were destroyed for aggressive shrimp culture. Shrimp farming also has increased salinity in the soil, reduced agricultural production; earnings of the peasants; livestock, and employment. Table 1 shows the total area (in hectares) of shrimp culture and production (in metric tons) in 2012-2013.

Despite all the negative impacts of shrimp farming on the environment, people are still employing cultivable land. Figure 1 shows the trends of areas of cultivated land and a yearly production of shrimp farming from FY 2008-2009 to FY 2012-2013 in Chittagong and Cox's Bazar districts. Plate (a) in figure 1 shows the trend of areas and production in the Chittagong district, and plate (b) shows the trends of areas and production for Cox's Bazar district. The plate (a) and (b) in figure 1 state an increasing trend for employing more cultivable land for shrimp farming though there was a fall in land employment in Cox's Bazar during FY2010-2011 to FY2011-2012.

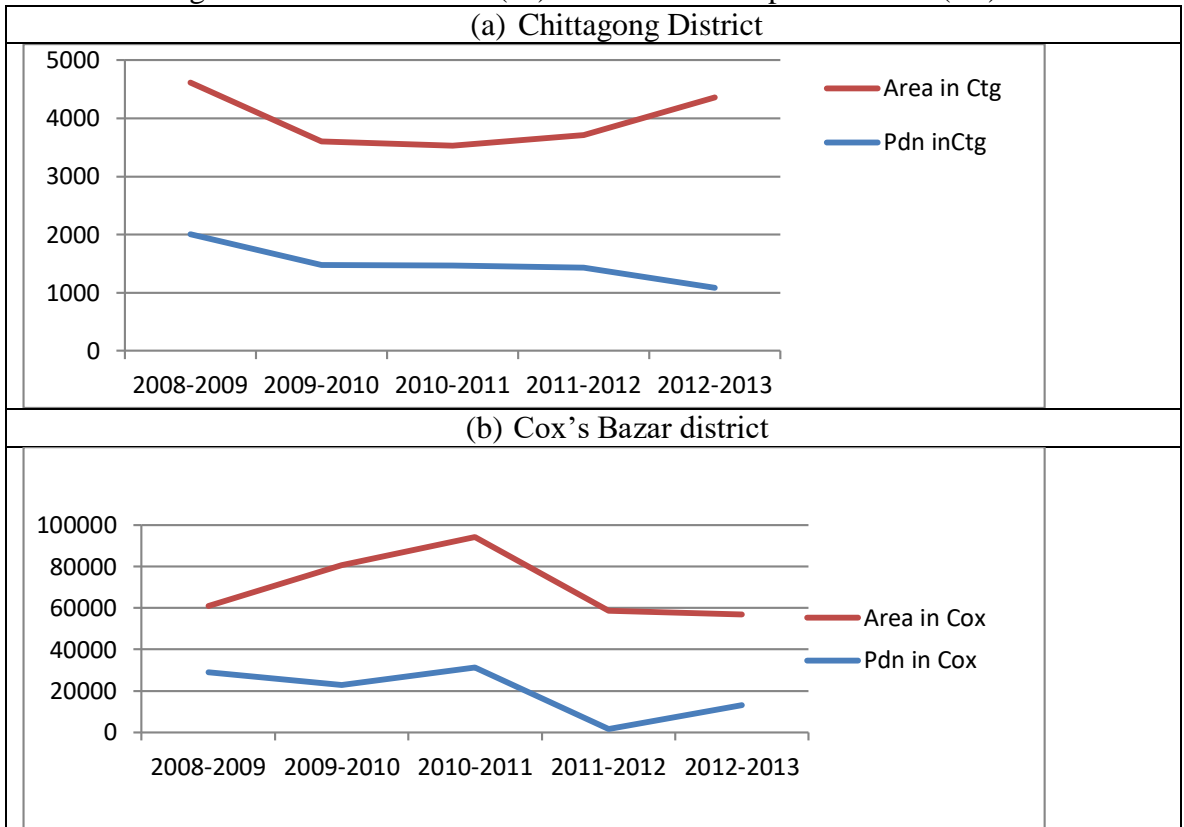

Figure 1: Cultivated area (ha) & Annual Shrimp Production (Mt).

*Source: Statistical Year Book (FY 2008-2009 to FY 2012-2013)*

This study evaluates the environmental cost of shrimp farming in selected coastal areas in Bangladesh and finds alternatives to the usual practice to minimize the cost of the environment. These objectives are served by examining previous articles and analyzing literature and relevant web pages to pinpoint the cost of shrimp production in the environment. In this relation, theories of environmental economics were considered along with the secondary research works to suggest the policies to reduce the environmental cost in the coastal areas so that the shrimp culture may develop into a sustainable aquaculture system.

2. **Study Area**

The coastal areas in Bangladesh are suitable for shrimp farming. As a result, two coastal districts in Chittagong Division are considered as the study area for this study. There is one upazila (Banshkhali) from Chittagong district and two upzila (Pekua and Moheshkhali) from Cox's Bazar district were selected for sampling based on the density of shrimp farming. There were 10 unions from these three upazila that have chosen for collecting sample data. The detail demographic information of the study areas are shown in Table 2.

Table 2: Demography of the Study Area

| Division | Chittagong | | | | | | | | | |
|---|---|---|---|---|---|---|---|---|---|---|
| District | Chittagong | | | Cox'sbazar | | | | | | |
| Upazila | Banshkhali (376.9 SKM) | | | Pekua | | | Moheshkhali (362.18) | | | |
| Union | Gandamara | Baghmara/ Katharia | Saral | Magnama | Rajakhali | Bara Bakia | Kutubjom | Boro Maheshkhali | Hoanak | Kalarmarchara |
| Area(Acres) | 7343 | 2710 | 7499 | 5302 | 4377 | 3459 | 7527 | 3706 | 9195 | 7174 |
| Households | 6456 | 3782 | 7185 | 3982 | 4551 | 3640 | 5367 | 8149 | 9373 | 8930 |
| Pop. | 33265 | 20866 | 38304 | 22088 | 25883 | 18430 | 30637 | 45068 | 51587 | 49268 |

*Source: BBS, 2012*

## 2.1. Banskhali Upazila

Banshkhali is the south eastern upazila of Chittagong district. Total area of Banskhali upazila is 376.9 skm (approximately 93135 acres), and the number of households is 84216 with the population of 431162. There are 14 union parishad in this upazila and among those three unions (Gandamara, Saral and Baghmara) are selected in this study for conducting the primary survey.

## 2.2 Pekua Upazila

Pekua is one of the five upazila of Cox;s Bazaar district. The total population in this upazila is 171538 with 31944 households covering the total area of 34500 acres (BBS, 2012). Three out of the six union parishad of this upazila has been selected for data collection based on the availability of shrimp culture. The selected union parishads are Magnama, Rajakhali and Bara Bakia.

## 2.3 Moheshkhali Upazila

Moheshkhali is another upazila in Cox's Bazaar district. The area of Moheshkhali Upazila is about 362 skm (approximately 89498 acres). The number of households in this upazila is 58177 with the population of 321218. This study has selected four union parishads of this upazila out of eight union parishad. The unions are Kutubjom, Boro Moheshkhali, Hoanak and Kalarmarchara.

**Map1: Location of the Study Area**

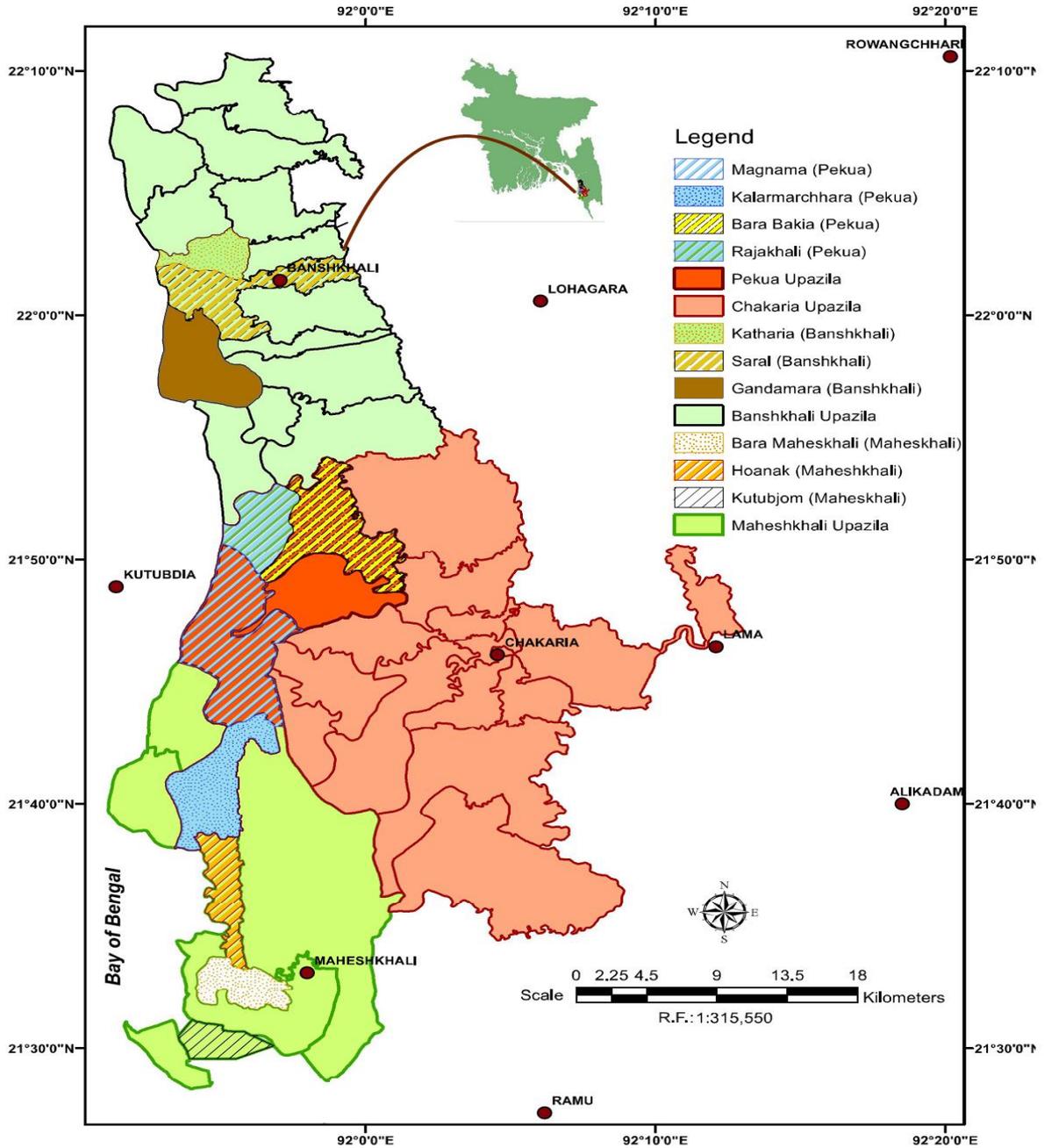

## 3. Literature Review

Shrimp culture has a great deal of environmental impact in the coastal areas, significantly influencing residents' environmental concerns (Alam and Hossain, 2018). Several previous articles and journals were studied to find the cost of shrimp farming on the environment. Some essences of those studies are scrutinized here.

Dieberg & Kiattisimul (1996) infers that shrimp culture destroys coastal zones and a sustainable aquaculture system in Thailand. They found that from 1979 to 1993, 16 to 32% of mangrove forest was destroyed from shrimp culture. That is why they suggested that implementing governmental policy on intensive pond culture requires the promotion of adequate training and awareness; education among farmers; monitoring and enforcing their activities; rehabilitating abandoned ponds; managing lands used in coastal zones coupled with community involvement and government reorganization to overcome the gaps.

Gunawardana and Rowan (2005) conducted a study on calculating the cost of shrimp production on the mangrove ecosystem in Sri Lanka. They have calculated that 24% of mangroves were harvested for firewood, 10% for fish or prawn, 8% for construction materials and other miscellaneous products as medicine, honey, tansies for dyes and manure. These all are the direct contribution of mangroves. The indirect values they considered are protection from waves and storms to the coastal area, saving lives and properties. They have used TEV (Total Environmental Value) to calculate the cost of shrimp farming on the mangrove ecosystem. In calculating the TEV, they added the direct use value, indirect use value, option value, and existence value of mangrove forests.

Primavera (1999) has emphasized mangrove clearance as a cause of life and property damage in the coastal area in the Philippines. Whereas for explaining the causes of mangrove loss, he has identified shrimp farming as one of the most crucial causes of mangrove destruction.

Billie (2000) summarized that coastal property right is quite complicated among the stakeholders in Mexico (i.e., federal zones, sub-divided ejido land, communal ejido lands, private property, and co-operatives). Thus, undefined property rights mainly contribute to the conflict among the shrimp farmers and the denizens of the coastal area.

Pa'ez-osuna (2001) discussed that an environmentally non-degrading, technically appropriate, economically viable, and socially acceptable shrimp farming practice refers to sustainable aquaculture. The socio-economic cost of modern shrimp farming was proven by a cost-benefit analysis (weighing the pros and cons) by Indian Supreme Court. The result shows that shrimp culture caused more economic harm than good as the damage outweighed benefit by 4 to 1 in Andra Prodesh (63 billion rupees vs. 15 billion rupees annual earnings) and 1.5 to 1 in Tamil Nadu followed by a destruction cost of mangrove, salinity and increasing unemployment (Primavera, 1997 ).

Umamaheswari et al. (2011) have examined the Salinisation in two villages in the Southern India. They have considered the salinisation induced by shrimp farming and concluded that one village Poovam is affected by salinisation while other village Thiruvettakudy is not. The soil sample they have taken at 2006 exhibits the presence of salinisation in Poovam though it was at the normal level during 1945-1995 shrimp farming. Thus in estimating the paddy production function they found that salinisation has negative and significant effect on paddy production. Their finding shows that reducing the saline level to normal level farm can gain Rs. 1000 to Rs. 5000 per hectare. Chowdhury et al. (1994) reported that coastal communities in Bangladesh are most vulnerable to tropical storms because of extensive mangrove clearance. The effects are visible by suffering with various cyclones and tidal waves that causes loss of thousands of lives and livestock in Bangladesh in every year. They also mentioned that 7500 hectares of mangrove (in 1975) was almost cleared and reduced to 973 hectares (in 1988) in Chokoria Sunderbans for shrimp farming. So, the wide scale expansion of shrimp culture has directly reduced mangrove sharply in the coastal part of Bangladesh. They have concluded that the Chakaria mangrove has protected the villagers from a tidal wave in 1969 but the devastating cyclone in 1991 has caused about 147,000 death and enormous damages of properties after the introduction of shrimp farm (CF: Primavera, 1997). The rapid expansion of shrimp farms also increase the level of salinity in the coastal cultivable lands and reduce the bio diversity in the coastal area (Toufiq, 2002).

The literature reviewed above showed that there is a gap in research which deals the environmental cost of shrimp farming as a whole in Bangladesh coastal areas. Though there are many research works which considers different services of ecosystem, but no significant studies found which deal the environmental costs of shrimp farming as a measure of foregone benefits in monetary values. This study thus estimates the environmental cost of shrimp farming in terms of forgone values of ecosystem services. But the focus is on the coastal area in Chittagong only, not to the whole coastal areas in Bangladesh.

### 4.1 Theoretical background

The aim of this study is to identify the environmental cost of shrimp farming in coastal area of Bangladesh. This cost is projected as the loss of total environmental value of the coastal ecosystem derived by the introduction or practice of shrimp farming in this area. That is the total environmental cost is considered as the loss of total value of the environmental services. Thus, in calculating the total environmental value (TEV) the monetary value of i) direct use value ii) indirect use value iii) option value and iv) existence value were summed up (Gunnawara & Rowan, 2005). As a result, the total environmental cost of shrimp farming stands for the environmental value lost associated with the shrimp culture. That is,

$$TEC(=TEV) = \text{Cost of direct use value} + \text{Cost of indirect use value} +$$
$$\text{Cost of existence value} + \text{Cost of option value} \quad\text{---------------} \quad (1)$$

Whereas the cost of direct use value involves forestry and fishery products that were possible to collect for households firewood support and loss of agricultural products harvested in the coastal area due to destruction of mangrove forest for shrimp production. The cost of indirect value for environment refers the loss of amenity services due to destruction of the mangrove forest in the coastal area. On the other hand, the cost of option value is the individuals willingness to pay (WTP) to preserve the resources for future uses and the cost of existence value is the individual's concern about the environment and the future use and existence of the resources (Alam, 2013; Perman et al. 2004). The total environmental cost thus stands for,

$$TEC = TEV = \sum_{i=1}^{n} MTWTP \quad\text{----------------------} \quad (2)$$
$$= MTWTPxNHhs$$

Here, NHhs stands for number of households, MTWTP stands for mean total willingness to pay. On the other hand, mean total willingness to pay is calculated as the sum of mean willingness to pay for four types of ecosystem services provided by the nature. That is,

$$MTWTP = \sum (MWTP_{DV} + MWTP_{IdV} + WTP_{OV} + MWTP_{EV}) \quad\text{----------} \quad (3)$$

Here, the mean willingness to pay for each category of ecosystem services is calculated as follows,

$$MWTP = \frac{\sum WTP}{n} \quad \text{---------------------------(4)}$$

Here n is refers the number of sample collected for each category of services. Finally, to calculate the environmental cost of shrimp farming, number of households for each unit of area or locality will be used from secondary sources (Table 2).

### 4.2 Data

The data for environmental cost of shrimp farming has been collected from both primary and secondary sources. The primary data has been collected using a questionnaire based random sampling from the coastal area in Cox's Bazar (two upazila) and Chittagong (one upazila) Districts. For collecting primary data, a set of three different questionnaires were prepared based on Contingent Valuation Method. The questionnaires were then used to collect information from different stakeholders like peasants, coastal households, coastal villagers and inhabitants of the coastal villages. Respondents were asked to state their willingness to pay for different services provided by the mangrove forest with a specified questionnaire and at the same time they were asked to assess the loss of their current profession or activity due to the presence of shrimp farming. In addition to this, few samples were collected from the owner of the shrimp farms to gather information in assessing the cost function for the shrimp farming. For collecting secondary data, Upazila statistics department, fishery department, Bangladesh Bureau of Statistics (BBS), and Statistical Year Books were used along with different published articles.

In primary data section, information collected from the coastal peasant's was considered for assessing the direct cost of shrimp farming. The people who are living close to the coastal belt receive the indirect support of mangrove forests by protecting their homes and wealth from flood and cyclone. Their information was considered as the indirect cost of shrimp farming since mangrove forest destroys due to increase of shrimp farming in the coast line. Finally, the information of the inhabitants of those areas was used to assess the cost of option value and the existence value. Detail sample plan has been shown in Table 3.

Table 3: Primary Data Sources with Sample Size

| District | Upazila | Sample size | Direct +Indirect +Option and Existence |
|---|---|---|---|
| Chittagong | Banshkhali | 69 | 23 * 3=69 |
| Cox's Bazar | Pekua | 138 | 46*3=138 |
| Cox's Bazar | Moheshkhali | 120 | 40 *3=120 |
| Total Sample | | 327 | = 327 |

## 4.3 Environmental costs of shrimp farming

Though shrimp farming has a role in earning foreign exchange, it has some negative external influences on the environment. Network for Aquaculture in Asia-Pacific (NACA), during the period of 2001-2010 reported that some forms of aquaculture (mainly shrimp and salmon) may be environmentally unsustainable, socially inequitable and are not safe for consumers (NACA web). Those adverse effects of shrimp farming can be treated as the cost on environment. The main effects of shrimp farming are destruction of mangrove forests, salinity of soil, losses of agricultural land and its productivity, migration and landlessness of the coastal community, social conflicts, and tidal waves etc.

### 4.3.1 Mangroves destruction

Mangrove provides many services like coastal stabilization, nursery ground for fish; crabs and shrimp, nutrient and sediment traps, BOD (Bio Oxygen Demand) removal; organic matter production and water consumption (Dieberg & Kiattismkul, 1996). Thus destruction of mangrove is very costly for the environment. Considering all services of mangrove, Dixon (1989; quoted from Dieberg & Kiattismkul, 1996) suggested that the annual value of mangrove is between USD 500 to USD 2500/ ha when fishery and forestry benefits are included. Whereas, Primavera (1993) noted the value of mangrove services is about USD 11,600/ ha / per year.

In November 2007 a tropical cyclone Sidr has destroyed one third area of Sundarban; the largest mangrove forest in the world which caused thousands of death of human lives (officially death were counted as 3,447 but Red Cross reported that death could be up to 10,000) and livestock, destroyed crops and houses of three big districts; Potuakhali, Barguna and Jhalokati of south-western part of Bangladesh. The effects of Sidr was comparatively less due to the intervention of the Sundarban mangrove forest though it was more destructive than 1991 cyclone, otherwise there were more damages in that part of Bangladesh. But mangroves started destroying after the installation of shrimp farming in the coastal area. Finding reported that 35% of global mangrove forests have been destroyed in between 1980 to 1990 (Valiela, 2001) after the installation of shrimp during 1970's. In addition to this, the effect of industrial shrimp farming is much higher since discharges nutrient rich effluent containing fertilizer, pesticide and antibiotic leading environmental pollution (Owen, 2004).

### 4.3.2 Salt water intrusion

Another effect of shrimp farming is salt water intrusion. Digging new ponds in the coastal area for shrimp farming affects the wetlands and rice growing areas by surface and sub-surface salt water intrusion. As a result, fresh water (in the agricultural area) becomes affected by saline water which reduces productivity of agriculture and thus the income of the peasant households. Monju (1996) found that household's income decreases due to decline in rice production, loss of poultry and livestock and erosion of homestead vegetation and social forestry. He also found that post-shrimp income of peasant household (local) was only 62% of their pre-shrimp culture level in Bangladesh. Thus, Thailand has banned inland shrimp farming (farming in fresh water zone) in 1999 due to much destruction of agricultural land by salinity (FAO, 2005).

### 4.3.3 Decrease in agricultural land and income

The expansion of unplanned and haphazard shrimp culture in coastal area has a direct effect on agricultural land. As shrimp culture is found profitable, the farmers use more cultivable land to dig ponds for shrimp culture which reduces agricultural feasibility in the coastal area. Hoque and Saifuzzaman (2002) agreed that unplanned and chaotic encroachment of rural agricultural land for shrimp industry has a noticeable impact on rural environmental condition in the coastal area. Karim (2006) shows that cropping intensity of rice decline only for shrimp cultivation in coastal area. He also found that wheat, jute and sugar can be affected and now it is not possible to grow for salinity in soil.

### 4.3.4 Decrease in livestock

In the last twenty years 64% of the lands used for cattle grazing decreased in the coastal areas in Bangladesh. The one crop lands which were used for grazing land for livestock are used for shrimp culture now. As a result, the number of livestock decreased in the coastal areas in Bangladesh. Care International, a prominent NGO studied that the average livestock per household in Bangladesh in 1980 was 2.3 which reduced to 1.6 in 1999 and the average number of goats per household declined from 3.3 in 1980 to 1.3 in 1999 (Anwar, 2003).

So, these indirect costs of shrimp farming on environment should be considered in the cost function to make it optimal. The shrimp farmers for their profit motives are not concerned about these environmental costs associated from their activities but for making shrimp farming socially optimal all external costs should be considered. The policy makers including the Department of Environment could think about these environmental costs for internalization in the production function of the shrimp farmers for comparing the monetary value of the benefits from this sector.

## 5   Result and Discussion

The result is shown in two different systems. Firstly, the Yes/No answers of the CVM from the questionnaire is shown in the following tables and discussed the individuals willingness to pay to avoid the environmental effects of shrimp farming by which environmental cost has been estimated.

### 5.1 Direct cost of shrimp farming

The following tables discuss individual's attitudes towards the environment. Table 4 shows that the almost 80% respondents are male while there is only 8% female respondents in the survey which is quite low.

Table 4: Knowledge about Environmental Problem

| Gender | Yes | No | Not Answered | Total |
|--------|-----|-----|--------------|-------|
| Male | 81 (74.3%) | 17 (15.5%) | 0 | 98 (89.9%) |
| Female | 7 (6.42%) | 4 (3.66%) | 0 | 11 (10.09%) |
| Total | 88 (80.73%) | 21 (19.26 %) | | 109 (100%) |

On the other hands, 63% male respondents replied (Table 5) that they are involved in shrimp farming while the rate is also very poor for female workers. Among the respondents about 60% respondents (Table 6) replied that they have the knowledge about the environmental costs of shrimp farming. But only, 9% respondents replied that they tried to do some sort of negotiation (Table 7) with the shrimp farmers though most of them (89%) informed that shrimp farming is available in their locality (Table 8).

Table 5: Involvement in shrimp farming

|  | Yes | No | Not Answered | Total |
| --- | --- | --- | --- | --- |
| Male | 69 (63.30%) | 29 (26.60 %) |  | 98 (89.9%) |
| Female | 1 (0.91%) | 10 (9.17%) |  | 11 (10.09%) |
| Total | 70 (64.22%) | 39 (35.77%) |  | 109 (100%) |

Among the respondents 76% replied that they have noticed the change in the production of agricultural goods (Table 5.6) though only 46% male (Table 5.7) wanted to remedy the effects showing their Willingness to pay. Interestingly, 81% respondents (Table 5.8) opined that they want the shrimp farms to be continued and only 8% of those respondents (Table 5.9) showed their willing to pay any kind of taxes or fees to avoid environmental cost.

Table 6: Knowledge about the environmental cost of shrimp farming

|  | Yes | No | Not Answered | Total |
| --- | --- | --- | --- | --- |
| Male | 58 (53.21%) | 40 (36.69%) | 0 (0%) | 98 (89.9%) |
| Female | 7 (6.42%) | 3 (2.75%) | 1 (0.91%) | 11 (10.09%) |
| Total | 65 (59.63%) | 43 ( 39.44) | 1 (0.91%) | 109 (100%) |

Table 7: Initiative to minimize the problem

|  | Yes | No | Not Answered | Total |
| --- | --- | --- | --- | --- |
| Male | 10 (9.17%) | 73 (66.97%) | 15 (13.761%) | 98 (89.908%) |
| Female | 0 (0%) | 7 (6.422%) | 4 (3.669%) | 11 (10.091%) |
| Total | 10 (9.174%) | 80(73.394%) | 19 (17.431%) | 109 (100%) |

Table 8: Presence of shrimp farms in respondent's locality

| Gender | Yes | No | Total |
|---|---|---|---|
| Male | 97 (88.99%) | 1 (0.91%) | 98 (89.908%) |
| Female | 11 (10.091%) | 0 (0%) | 11 (10.091%) |
| Total | 108 (99.082%) | 1 (0.917%) | 109 (100%) |

Though only 9% respondents in Table 7 has replied that they are concerned about the environmental costs of shrimp farming but 83% of them have replied that they have noticed the changes in their productions (Table 9). In connection to this change noticed in the production they want to avoid the adverse effects of shrimp farming through any policies under taken by the government. To pursue the policies the respondents have shown their willingness to pay if the government asks them to contribute for the avoidance of adverse effects of shrimp farming and it is about 49% (Table 10). Though the respondents wanted to avoid the adverse effects of shrimp farming but 73% of them (Table 11) allow the shrimp farming to be continued in their localities. This has been also proven by only 8% respondents (Table 12) reply of paying taxes to stop the shrimp farming in their areas.

Table 9: Noticed Changes in Production

|  | Yes | No | Not Answered | Total |
|---|---|---|---|---|
| Male | 83 (76.146%) | 13 (11.926%) | 2 (1.834%) | 98 (89.908%) |
| Female | 8 (7.339%) | 3 (2.752%) | 0 (0%) | 11 (10.091%) |
| Total | 91 (83.486%) | 16 (14.678%) | 2 (1.834%) | 109 (100%) |

Table 10: Have WTP to avoid the adverse effects

|  | Yes | No | Not Answered | Total |
|---|---|---|---|---|
| Male | 50 (45.871%) | 33 (30.375%) | 15 (13.761%) | 98 (89.908%) |
| Female | 3 (2.752%) | 4 (3.669%) | 4 (3.669%) | 11 (10.091%) |
| Total | 53 (48.623%) | 37 (33.944%) | 19 (17.4311%) | 109 (100%) |

Table 11: Allows shrimp farming to be continued

|  | Yes | No | Not Answered | Total |
|---|---|---|---|---|
| Male | 77 (70.642%) | 19 (17.431%) | 2 (1.834%) | 98 (89.908%) |
| Female | 4 (3.669%) | 6 (5.504%) | 1 (0.917%) | 11 (10.091%) |
| Total | 81 (74.311%) | 25 (22.935%) | 3 (2.752%) | 109 (100%) |

Table 12: Willingness to pay taxes to stop shrimp farming

|  | Yes | No | Not Answered | Total |
|---|---|---|---|---|
| Male | 7 (6.422%) | 15 (13.761%) | 76 (69.724%) | 98 (89.908%) |
| Female | 2 (1.824%) | 4 (3.66%) | 5 (4.587%) | 11(10.091%) |
| Total | 9 (8.256%) | 19 (17.431%) | 81 (74.311%) | 109 (100%) |

## 5.2 Indirect cost of shrimp farming (loss of mangrove forest)

The indirect cost of shrimp farming is considered as the loss of mangrove forest in the coastal area which can protect lives and assets from the damages of floods and cyclones.

Tables 13-20 are associated with the survey findings for indirect cost of shrimp farming. Respondents were asked about the benefits of mangrove forests and surprisingly 96% of them replied (Table 13) that they are aware of the benefits of mangrove on their lives. Those who are aware of the benefits of mangroves also replied (59%) that it helps saving the properties from the storms and floods (Table 14).

Table 13: Knowledge about benefits of mangrove

|  | Yes | No | Not Answered | Total |
|---|---|---|---|---|
| Male | 94 (86.238%) | 4 (3.669%) | 1 (0.917%) | 98 (89.908%) |
| Female | 11 (10.091%) | 0 (0%) | 0 (0%) | 11 (10.091%) |
| Total | 105 (96.33%) | 3 (2.752%) | 1 (0.917%) | 109 (100%) |

Table 14: Mangrove saving properties from floods and storms

|        | Yes            | No             | Not Answered | Total         |
|--------|----------------|----------------|--------------|---------------|
| Male   | 56 (51.376%)   | 37 (33.944%)   | 5 (4.587%)   | 98 (89.908%)  |
| Female | 6 (5.504%)     | 5 (4.587%)     | 0 (0%)       | 11 (10.091%)  |
| Total  | 62 (56.88%)    | 42 (38.532%)   | 5 (4.59%)    | 109 (100%)    |

In this connection 96% of the respondents (Table 15) expect that the mangrove exists for the future. To fulfill their desire of existence of the mangrove 72% respondents (Table 16) showed their willingness to pay.

Table 15: Desire for existence of Mangrove forest

|        | Yes            | No           | Not Answered | Total         |
|--------|----------------|--------------|--------------|---------------|
| Male   | 94 (86.24%)    | 1 (0.917%)   | 3 (2.752%)   | 98 (89.908%)  |
| Female | 11 (10.091%)   | 0 (0%)       | 0 (0%)       | 11 (10.091%)  |
| Total  | 105 (96.33%)   | 1 (0.917%)   | 3 (2.752%)   | 109 (100%)    |

Table 16: WTP to protect the mangrove forest for existence

|        | Yes            | No             | Not Answered | Total         |
|--------|----------------|----------------|--------------|---------------|
| Male   | 72 (66.06%)    | 24 (22.018%)   | 2 (1.834%)   | 98 (89.908%)  |
| Female | 6 (5.504%)     | 5 (4.58%)      | 0 (0%)       | 11 (10.091%)  |
| Total  | 78 (71.56%)    | 29 (26.605%)   | 2 (1.834%)   | 109 (100%)    |

Table 17: Knowledge about the shrimp farms in their locality

|        | Yes            | No             | Not Answered | Total         |
|--------|----------------|----------------|--------------|---------------|
| Male   | 81 (74.31%)    | 16 (14.685)    | 1 (0.917%)   | 98 (89.908%)  |
| Female | 5 (4.58%)      | 6 (5.504%)     | 0 (0%)       | 11 (10.091%)  |
| Total  | 86 (78.89%)    | 22 (20.183%)   | 1 (0.917%)   | 109 (100%)    |

The finding also shows that (Table 17) almost 79% respondents know the number of shrimp farms in their locality and 54% of them (Table 18) noticed the changes in the mangrove forest after the installation of the shrimp farms in their villages or unions.

Table 18: Noticed changes in mangrove forest after shrimp farming

|  | Yes | No | Not Answered | Total |
| --- | --- | --- | --- | --- |
| Male | 54 (49.541%) | 41 (37.61%) | 3 (2.75%) | 98 (89.908%) |
| Female | 5 (4.58%) | 6 (5.504%) | 0 (0%) | 11 (10.091%) |
| Total | 59 (54.128%) | 47 (43.119%) | 3 (2.75%) | 109 (100%) |

Table 19: Desire to protect mangrove forest from destruction

|  | Yes | No | Not Answered | Total |
| --- | --- | --- | --- | --- |
| Male | 92 (84.40%) | 4 (3.669%) | 2 (1.834%) | 98 (89.908%) |
| Female | 11 (10.091%) | 0 (0%) | 0 (0%) | 11 (10.091%) |
| Total | 103 (94.495%) | 4 (3.669%) | 2 (1.834%) | 109 (100%) |

Table 20: Showed WTP to protect mangrove forest

|  | Yes | No | Not Answered | Total |
| --- | --- | --- | --- | --- |
| Male | 68 (62.39%) | 21 (19.266%) | 9 (8.256%) | 98 (89.908%) |
| Female | 5 (4.587%) | 6 (5.505%) | 0 (0%) | 11 (10.091%) |
| Total | 73 (66.972%) | 27 (24.77%) | 9 (8.256%) | 109 (100%) |

In protecting the mangrove forest 67% of the respondents (Table 20) showed their willingness to pay if there is any demand for it. Though some of them did not notice mangrove destruction but 94% of the respondents (Table 19) want to protect the mangrove from the destruction.

## 5.3 Cost of option value and existence Value

Tables 21-27 summarized the findings of cost of option and existence values from the survey data. In calculating the option cost and existence cost, the consideration was only the option value and existence value of mangrove forest lost due to the shrimp farming. In asking questions to the respondents, a scenario has been described on shrimp farming which created a threat to coastal mangrove forest which has the option value and existence value.

Table 21 shows that 90% respondents have the knowledge about the importance of mangrove forest. Mangrove forest in the coastal area has a tourist value as well and 62% of the respondents (Table 22) have replied that they have visited the mangrove forest as a place of tourism.

Table 21: Knowledge about the importance of mangrove forest

|  | Yes | No | Not Answered | Total |
| --- | --- | --- | --- | --- |
| Male | 88 (83.81%) | 8 (7.62%) | 0 (0%) | 96 (91.43%) |
| Female | 7 (6.67%) | 2 (1.904%) | 0 (0%) | 9 (8.57%) |
| Total | 95 (90. 48%) | 10 (9.523%) | 0 (0%) | 105 (100%) |

Table 22: Visited mangrove forest

|  | Yes | No | Not Answered | Total |
| --- | --- | --- | --- | --- |
| Male | 64 (60.95%) | 32 (30.48%) |  | 96 (91.43%) |
| Female | 1 (0.952%) | 8 (7.62%) |  | 9 (8.57%) |
| Total | 65 (61.904%) | 40 (38.09%) |  | 105 (100%) |

Table 23: Have a plan to visit Mangrove forest soon

|  | Yes | No | Not Answered | Total |
| --- | --- | --- | --- | --- |
| Male | 39 (37.14%) | 56 (53.33%) | 1 (0.95%) | 96(91.43%) |
| Female | 1 (0.95%) | 8 (7.625) | 0 (0%) | 9 (8.57%) |
| Total | 40 (38.09%) | 64 (60.95%) | 1 (0.95%) | 105(100%) |

Table 24: Have a plan to visit mangrove in future

|  | Yes | No | Not Answered | Total |
|---|---|---|---|---|
| Male | 37 (35.23%) | 24 (22.86%) | 35 (33.33%) | 96(91.43%) |
| Female | 1 (0.95%) | 7 (6.67%) | 1 (0.95%) | 9 (8.57%) |
| Total | 38 (36.19%) | 31 (29.52%) | 36 (34.29%) | 105(100%) |

There are 38% respondents (Table 23) who replied that they have a plan to visit the mangrove forest very soon though 36% of the respondents (Table 24) replied that they will visit the mangrove forest in the future. In connection to their visit plan, 32% respondents (Table 25) showed their willingness to pay for the conservation for the mangrove forest

Table 25: Have WTP for the conservation of mangrove forest

|  | Yes | No | Not Answered | Total |
|---|---|---|---|---|
| Male | 32 (30.48%) | 10 (9.52%) | 54 (51.42%) | 96(91.43%) |
| Female | 2 (1.90%) | 1 (0.95%) | 6 (5.71%) | 9 (8.57%) |
| Total | 34 (32.38%) | 11 (10.48%) | 60 (57.14%) | 105(100%) |

Table 26: Desire for existence of mangrove forest

|  | Yes | No | Not Answered | Total |
|---|---|---|---|---|
| Male | 62 (59.04%) | 0 (0%) | 34 (32.38%) | 96(91.43%) |
| Female | 6(5.71%) | 1 (0.95%) | 2 (1.905) | 9 (8.57%) |
| Total | 68 (64.76%) | 1 (0.95%) | 36 (43.28%) | 105(100%) |

Table 27: WTP for the conservation of mangrove forest

|  | Yes | No | Not Answered | Total |
|---|---|---|---|---|
| Male | 66 (62.865) | 18 (17.14%) | 12 (11.42%) | 96(91.43%) |
| Female | 3 (2.86%) | 2 (1.90%) | 4 (3.81%) | 9 (8.57%) |
| Total | 69 (65.71%) | 20 (19.04%) | 16 (15.23%) | 105(100%) |

For the existence of the mangrove, 65% of the respondents (Table 26) showed a positive desire whereas, 66% respondents (Table 27) showed a willingness to pay for the conservation of the forest either for existence or option values it creates for us.

## 6  Empirical Findings

After the data compilation mean willingness to pay is calsculated by dividing the total value of each category by the total number of respondents. Then all mean willingness to pay is added to get the mean total willingness to pay as a proxy of per acre of mean total environmental cost of shrimp farming. The mean total environmental cost of shrimp farming then transformed into USD to make it more convenient for comparison with other countries. In addition to this median willingness to pay is also shown in the following table. It is found that both mean willingness to pay and median willingness to pay is calculated but mean willingness to pay is widely used in most of the estimations (Perman et al. 2004; pp. 425).

Table 28: Calculating Mean WTP (Mean Cost) Per acre/year

|  | Direct Cost in BDT | Indirect Cost in BDT | Existence cost in BDT | Option cost in BDT | Mean/Median Total in BDT | Total in USD |
|---|---|---|---|---|---|---|
| Mean WTP/Acre | 285.6 | 339.71 | 187.33 | 228.48 | 1066.64 | $ 13.67 |
| Median WTP/Acre | 100 | 200 | 200 | 120 | 620 | $7.95 |

*Note 1USD= BDT 78*

Now considering the mean total cost of shrimp farming (mean total willingness to pay) from the above table the total environmental cost of shrimp farming has been calculated in the following table (Table 6.2). For the simplicity the mean total willingness to pay as environmental cost of shrimp is multiplied by the number of households (Table 2) in each level of area. The rest of the information was collected from BBS Census 2011.

Table 29: Total Environmental Cost of Shrimp Farming

| Union Parishad | | | Union Parishad | | | Union Parishad | | | |
|---|---|---|---|---|---|---|---|---|---|
| Gandamara | Baghmara/ Katharia | Saral | Magnama | Rajakhali | Bara Bakia | Kutubjom | Boro Maheshkhali | Hoanak | Kalarmarchara |
| 6886227.84 ($88284.97) | 4034032.48 ($51718.37) | 7663808.4 ($98253.95) | 4247360.48 ($54453.34) | 4857478.56 ($62275.37) | 3882569.6 ($49776.5) | 5724656.88 ($73393.04) | 8692049.36 ($111436.53) | 9997616.72 ($128174.57) | 9525095.2 ($122116.61) |
| Banshkhali Upazila | | | Pekua Upazila | | | Moheshkhali Upazila | | | |
| 89828154.24 ($1.152 M) | | | 34072748.16 ($4.37) | | | 62053915.28 ($7.955 M) | | | |
| Chittagong District | | | Cox's bazar District | | | | | | |
| 1634107413.00 ($20.95 M) | | | 443673174.6 ($ 5. 688) | | | | | | |

*Note: Households information was collected from BBS Census 2011*

The calculated results (Table 29) for environmental cost of shrimp are shown for union level, upazila level and later on district level. For ethical ground of view divisional level environmental cost is not calculated since shrimp farming and coastal area is not available in all the districts in Chittagong division. The result shows that yearly environmental cost of shrimp in Banskhali Upazila is $1.152 M where is it it $4.37 M and $7.955 M in Pekua Upazila and Moheshkhali Upazila respectively. It is clearly seen that environmental cost is very high in the upazila's in Cox'sbazar district comparing to a upazila in Chittagong district. It is quite likely and expected as there are many shrimp culture in different upazilas in Cox'sbazar distruct. In contrast, district level environmental cost is very high in Chittagong comparing to Cox'sbazar district. One probable cause of this dissimilar result might be for the higher number of households in Chittagong division.

## 7  Recommendations to Recovery the Environmental Costs

Shrimp culture has been profitable in Bangladesh as it earns foreign exchange; generates employments and uses natural resources. Though it has some negative effects on environment in the coastal areas but with proper policies, monitoring and enforcement of law and order can help it to remain a sustainable aquaculture. In view of this, some recommendations are proposed and discussed to make the shrimp farming a sustainable and profitable occupation.

## 7.1 Property rights and willingness to pay

Property right is an important tool to reduce the negative externality of an activity, especially for the environmental services as property right is poorly defined here. If property right is well defined for shrimp farming, the environmental cost of shrimp firming could be minimized with the help of Coasian-theorem (1960) since it satisfies the Coasian condition of costs outweighing the polluter's returns. For example, say the property right goes for shrimp farmers, and then he/she will have the right to produce. In this situation the coastal community may come to a negotiation and will be willing to pay the shrimp farmers not to produce shrimp in a large scale. By this negotiation community can reduce the destruction of mangrove forests, save agricultural land from salinity and unproductive. It will also lower the risk of damages from natural disasters like tidal waves and cyclones. Assessing all the costs of shrimp farming the coastal community will show their desire of how much to pay (WTP) not to produce shrimp in unplanned and hazardous way.

Figure 2: Optimal solution using property right

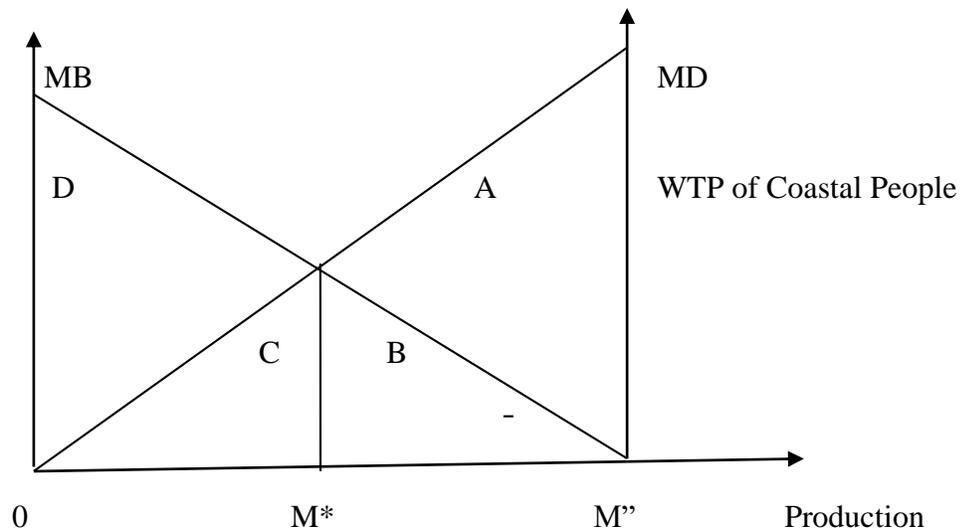

On the contrary, if the property right is defined in favor of coastal area peoples, then the shrimp farmers will have no right to impose negative externality on the environment which makes the coastal peoples worse off. Thus to continue their present activities, the shrimp farmers will be asked to pay (WTP) to the coastal peoples so that their (coastal people's) cost becomes minimum. This negotiation could be explained by the help of Figure 2.

For simplicity say, we consider that the property right is defined in favor of costal area peoples. Thus the shrimp farmers will pay to the community for creating damage to them because of shrimp farmer's activities. If they produce M* unit of shrimp, it creates marginal benefit equivalent to sum of area D and C. On the contrary, producing till M* creates marginal damage for the coastal people equivalent to area C. Thus to maintain M* unit of shrimp production the shrimp farmers will pay area C to the coastal people as compensation. This study suggests that the marginal damage of shrimp farming as a compensation is equivalent to $13.66 per acre per year (Table 28).

## 7.2 Permit/ license

One of the best solutions to control shrimp farming could be introducing Permit or Licensing. Permit can help sustainable shrimp farming as well as reduce destruction of other marine species while collecting wild fry. In Maxico (in Sinaloa) 3 companies are given Permits to collect PL (Post- Larva) around the bay of Ceuta. Some of the collected PL's were used for their own production and rests of them were sold to other farms. Other cooperatives were given Permits in San Blas in Nyarit. As a result, issuing permits for collecting PL for shrimp culture saved the unnecessary marine species which were destroyed before. The study suggests that Bangladesh can introduce the permit pricing which may be helpful for marine ecology and sustainable shrimp farming.

## 7.3 Tax on Shrimp production

Another alternative of controlling shrimp farming could be environmental tax on shrimp production. Though Shrimp has been treated as an export item and government has been given fiscal incentives to shrimp culture as it created more employment opportunities in coastal areas and earns foreign exchange, but considering its negative externalities on environment the adjustment policy can be taken to make it a sustainable aquaculture. If environmental tax is imposed then production will decrease and price of shrimp will increase. The rise of shrimp price may inversely affect exports of Bangladesh shrimp. But, it will also raise a moral question, how much the importing countries (USA, Canada, Japan and EU countries) are willing to pay for shrimp as a green product. If the importing countries go green, it will not be a problem at all. But, if they don't, then Bangladesh may lose some portion of market for shrimp export.

**7.4 Best Management Practices (BMPs)**

According to NACA (Network of Aquaculture Centers in Asia-Pacific) Best Management Practices can be used to make shrimp culture sustainable. They found that well designed and implemented BMPs help to increase efficiency and productivity by reducing shrimp health problems, reducing or mitigating the impacts of farming on environment, improving food safety and quality and improving the social benefits from farming and social acceptability and sustainability. The BMPs are used in India and Vietnam (NACA, Webpage). Bangladesh may use these management practices so that the shrimp farming becomes a sustainable aquaculture.

**8. Limitations and Conclusion**

Shrimp farming has been reported as an important economic activity all over the world in the last thirty years despite its negative externality on environment, in particular coastal ecosystems in the farming areas. The aim of this study was to estimate cost of that environmental degradation created by shrimp culture in the coastal area in Bangladesh. In this regard, primary data has been collected using a structured questionnaire. Though this total number of sample size is 327 it is comparatively small in each category of the Contingent Valuation Approaches. In addition to this, the people in the coastal area are not aware of the approaches used in this study. The limitation that the study faced is to consider the family size or the population size in calculating the total cost through the total environmental value. So, considering all limitations, this study concludes that we should take necessary initiatives to make the shrimp farming as an optimal, sustainable and an environmental friendly means of aquaculture since it is the third export earnings item in Bangladesh, contributes 9% of national exports (Hoque, 1999) enhances employment and uses of natural resources . Thus to internalize the environmental costs associated with shrimp farming the government of Bangladesh could encourage proper management practices, impose new policies (like property right or permit) and strengthen the monitoring activities so that this sector comes environment friendly means of production which may also play an important role in contributing the economy.